\documentclass[%
 aip,
 cha,%
 amsmath,amssymb,
 reprint,%
]{revtex4-1}

\usepackage{hyperref}
\usepackage{graphicx}% Include figure files
\usepackage{dcolumn}% Align table columns on decimal point
\usepackage{bm}% bold math
\usepackage{color,todonotes}
\usepackage[utf8]{inputenc}

\begin{document}

%\preprint{AIP/123-QED}

% Title
\title[]{Dimension-scalable recurrence threshold estimation}

% Authors
\author{K.~Hauke Kr\"amer}
\email{hkraemer@pik-potsdam.de, hkraemer@uni-potsdam.de}
\affiliation{Potsdam Institute for Climate Impact Research, Telegrafenberg A31, 14473 Potsdam, 
Germany, EU
}%
\affiliation{%
Institute of Earth and Environmental Science, University of Potsdam, Karl-Liebknecht-Str. 
24-25, 14476 Potsdam-Golm, Germany, EU
}%

\author{Reik V. Donner}%
\affiliation{Potsdam Institute for Climate Impact Research, Telegrafenberg A31, 14473 Potsdam, 
Germany, EU
}%

\author{Jobst Heitzig}%
\affiliation{Potsdam Institute for Climate Impact Research, Telegrafenberg A31, 14473 Potsdam, 
Germany, EU
}%

\author{Norbert Marwan}
\affiliation{Potsdam Institute for Climate Impact Research, Telegrafenberg A31, 14473 Potsdam, 
Germany, EU
}%

\date{\today}

% Abstract
\begin{abstract}
The appropriate selection of recurrence thresholds is a key problem in applications of recurrence quantification analysis (RQA) and related methods across disciplines. Here, we discuss the distribution of pairwise distances between state vectors in the studied system's state space reconstructed by means of time-delay embedding as the key characteristic that should guide the corresponding choice for obtaining an adequate resolution of a recurrence plot. Specifically, we present an empirical description of the distance distribution, focusing on characteristic changes of its shape with increasing embedding dimension. Based on our results, we recommend selecting the recurrence threshold according to a fixed quantile of this distribution. We highlight the advantages of this strategy over other previously suggested approaches by discussing the performance of selected RQA measures in detecting chaos-chaos transitions in some prototypical model system.
\end{abstract}


\maketitle

% lead paragraph
\begin{quotation}
Recurrence plots provide an intuitive tool for visualizing the (potentially multi-dimensional) trajectory of a dynamical system in state space. In many applications, however, only univariate (single-variable) observations of the system's overall state are available. In such cases, qualitatively reconstructing the action of unobserved components by means of embedding techniques has become a standard procedure in nonlinear time series analysis. In addition to the appropriate choice of embedding technique and parameters, the qualitative and quantitative recurrence properties of the reconstructed higher-dimensional time series depend on the selection of some specific norm to define distances in the corresponding metric space and an associated threshold distance $\varepsilon$ to distinguish close from distant pairs of state vectors. While the impact of $\varepsilon$ on RQA and related techniques has already been studied, the associated interplay with the embedding dimension has not yet been explicitly addressed. In turn, automatically and consistently selecting embedding parameters and recurrence threshold is key 
to make recurrence analysis generally applicable to researchers from a broad range of scientific disciplines. Here, we discuss a strategy for threshold selection that makes the results of RQA widely independent of the (sufficiently high) embedding dimension.
\end{quotation}

% Section 1 , Introduction

\section{Introduction}\label{introduction}

A vector time series $\{\vec{x}_i\}_{i=1}^N$ (with $\vec{x}_i=\vec{x}(t_i)$) 
provides an approximation 
of a specific trajectory of a given dynamical system in finite-time and (for time-continuous 
dynamical systems) finite-resolution. In many real-world applications, however, inferring 
complete dynamical information from observations is hampered by the fact that only some of the 
dynamically relevant variables are directly observable. In 
such cases, it has been demonstrated\cite{Tak} that it is possible to qualitatively reconstruct 
representations of the unobserved components of a higher-dimensional system by means of 
embedding techniques applied to a suitably chosen individual component\cite{Letellier2002}. 
Specifically, time-delay embedding has become a widely utilized method in nonlinear time series 
analysis, where a series of univariate observations $\{x_i\}$ (the actual time series at hand) is unfolded into a sequence of 
$m$-dimensional state vectors $\{\vec{x}_i\}$\cite{Tak,Pac} defined as $\vec{x}_i=(x_i,x_{i-
\tau},\dots,x_{i-(m-1)\tau})^T$, where $m$ and $\tau$ denote the chosen embedding dimension and 
embedding delay, respectively.

Introduced by Eckmann et al.\cite{Eckmann}, recurrence plots (RPs) provide a versatile tool 
for visualizing and quantitatively analyzing the succession of dynamically similar states in a 
time series. For this purpose, dynamical similarity is measured in terms of some metric 
distance $d_{i,j}=\|\vec{x}_i-\vec{x}_j\|$ defined in the underlying system's (reconstructed) 
state space. Based on the resulting distance matrix $\mathbf{d}=(d_{i,j})$, a recurrence matrix 
$\mathbf{R}=(R_{i,j})$ is defined as a thresholded version such that its entries assume values of 1 if the distance between the two associated state vectors is smaller than a threshold $
\varepsilon$, and 0 otherwise:
\begin{equation}
R_{i,j}(\varepsilon)=\begin{cases}1:&d_{i,j} \leq \varepsilon \\
0:&d_{i,j} > \varepsilon,\end{cases} \qquad i,j = 1,...,N.
\end{equation}
Equivalently, we can write
\begin{equation}
R_{i,j}(\varepsilon)=\Theta (\varepsilon - d_{i,j}), \qquad i,j = 1,...,N,
\end{equation}
where $\Theta(\cdot)$ is the Heaviside function. In this definition, the threshold $\varepsilon$ is 
fixed with respect to all pairwise distances contained in $\mathbf{d}$, and we will focus only on 
this kind of threshold application throughout this paper. An alternative definition of the recurrence 
matrix\cite{Eckmann,Marwan1}, which shall not be further considered in this study, replaces the 
global, fixed recurrence threshold $\varepsilon$ applied to all state vectors $\vec{x}_i$ by an adaptive 
local one that is defined such that the number of recurrences (i.e., close state vectors) is 
the same for each $\vec{x}_i$ (fixed amount of nearest neighbors)\cite{Eckmann}, leading to a constant local 
recurrence rate.

According to the above definition, for a given time series the recurrence matrix depends on the chosen recurrence threshold $\varepsilon$ together with the selected norm $\|
\cdot\|$ used for defining pairwise distances between the state vectors. In this work, we will 
restrict ourselves to two of the most commonly used norms: the Euclidean ($L_2$) and maximum ($L_\infty$, supremum, Chebychev) norms. Specifically, we will study how the distributions of 
pairwise $L_2$ and $L_\infty$ distances
depend on the embedding dimension.

Previous studies have provided various complementary suggestions for (i) selecting the right method of determining the recurrence threshold (i.e., a fixed or an adaptive approach) and (ii) choosing its actual value in some automatic way based on the specific properties of the system under study. Corresponding approaches include the spatial extent of the trajectory in the (reconstructed) state space\cite{Mind,Zbi1}, signal to noise 
ratio\cite{Zbi1,Zbi2,Thi1,schinkel2008}, the specific dynamical system underlying the time series under 
investigation\cite{Zbi2,Mat}, or properties of the associated recurrence 
network\cite{Marwan3,Don} with adjacency matrix $A_{i,j}=R_{i,j}-\delta_{i,j}$ (with $
\delta_{i,j}$ being the Kronecker symbol) like the percolation
threshold\cite{Donges2012,Jacob2016PRE}, 
second smallest eigenvalue of the graph's Laplacian\cite{Ero}, breakdown of $\varepsilon^{-1}$ 
scaling of the average path length\cite{Donges2012}, or information-theoretic 
characteristics\cite{Wie}. In practice, the appropriate choice of the method for determining the recurrence threshold, as well as its resulting value itself, can depend on the specific problem under study and take any of the above criteria or even some multiple-objective considerations based on different criteria into account. To this end, a general solution to the second problem of selecting a specific value of $\varepsilon$ has not yet been obtained, and we will also not address this problem specifically in the course of the present paper. Instead, we are attempting to provide some further insights into the first, more conceptual problem setting (i.e., which type of approach for selecting recurrence thresholds should be taken in case of varying situations such as different embedding dimensions).

As we will further detail in the course of this paper, the previously suggested approaches\cite{Koe,Mind,Zbi1,schinkel2008} to link a recurrence threshold to
a certain percentage of the maximal or mean distance of all pairwise distances 
of state vectors (i.e., a given fraction of the attractor's diameter in the reconstructed state space) causes the resulting recurrence characteristics not to be invariant with 
increasing embedding dimension. The reason for this behavior is as follows. In addition to a 
general increase of distances\cite{Zimek} (depending on the chosen norm)\cite{Koe}, the shape of the 
distance distribution also changes with increasing embedding dimension. This can result in a 
poorly resolved, almost white, RP and meaningless RQA characteristics mostly noticeable at higher embedding dimensions \textit{m}.

However, embedding a time series with $\textit{m}\sim \mathcal{O}(10^1)$ or even larger can become necessary when the correlation dimension $D_2$ of the attractor is rather 
large. This is due to the fact that Takens' theorem (and several extensions thereof) guarantee 
the existence of a diffeomorphism between the original and the reconstructed attractor if 
\textit{m} satisfies $\textit{m}\geqslant 2 D_2 +1$\cite{Marwan1,Tak,Sau}. Hegger et al.
\cite{Heg} emphasize that it is also advisable to choose a rather high value of $m$ when 
dealing with time series originating from a $D$-dimensional deterministic system that is 
driven by $P$ slowly time dependent parameters. An appropriate choice for $m$ then fulfills $m 
\geqslant 2(D+P)$. Concerning practical applications of nonlinear time series analysis in, e.g.,  Earth or Health science, one commonly deals with signals originating from complex, non-stationary systems and, therefore, high embedding 
dimensions are often necessary, requiring threshold selection methods which lead to robust results of RQA and related state space based techniques that are independent of the embedding dimension.

In the following Section~\ref{influence}, we study the influence of an increasing embedding 
dimension on the shape of the distance distribution in more detail. We deduce that, in order to 
avoid problems arising due to an unfavorable fixed recurrence threshold when varying $m$, 
we could choose $\varepsilon$ as a certain percentile of the distance distribution rather than a 
certain percentage of the maximum or mean phase space diameter. Successively, 
Section~\ref{numerical} presents a comparative study of some recurrence characteristics for the well-known
R\"ossler system in a setting with some time-dependent control parameter, highlighting the advantages of the proposed threshold estimation method. The 
main results of this study are summarized in Section~\ref{conclusion}.

% Section 2

\section{Influence of embedding dimension on the distance distribution}\label{influence}

In the following, we consider time series $\{x_i\}$ of length $N$ with a cumulative distribution function $P(x)$. As an overarching question, we study the variations in the maximum and 
mean pairwise distances of all pairs of state vectors when increasing \textit{m} from unity to 
some arbitrary maximum \textit{$m_f$}. In addition to $P(x)$ and $m$, the distribution of 
distances is expected to depend on the chosen norm used for the calculation of distances. Note 
that the effective number of state vectors $N_\text{eff}(m)=N-(m-1)\tau$ available for estimating 
the probability distribution of distances in $m$ dimensions will decrease with $m$. In order to 
avoid sample size effects in comparing the results for different $m$, we therefore choose $N$ 
sufficiently large so that $1-N_\text{eff}(m_f)/N\ll 1$.

% Section 2A

\subsection{Maximum norm}\label{influence_maximum}

\begin{figure}
 \centering
 \includegraphics[scale=0.25]{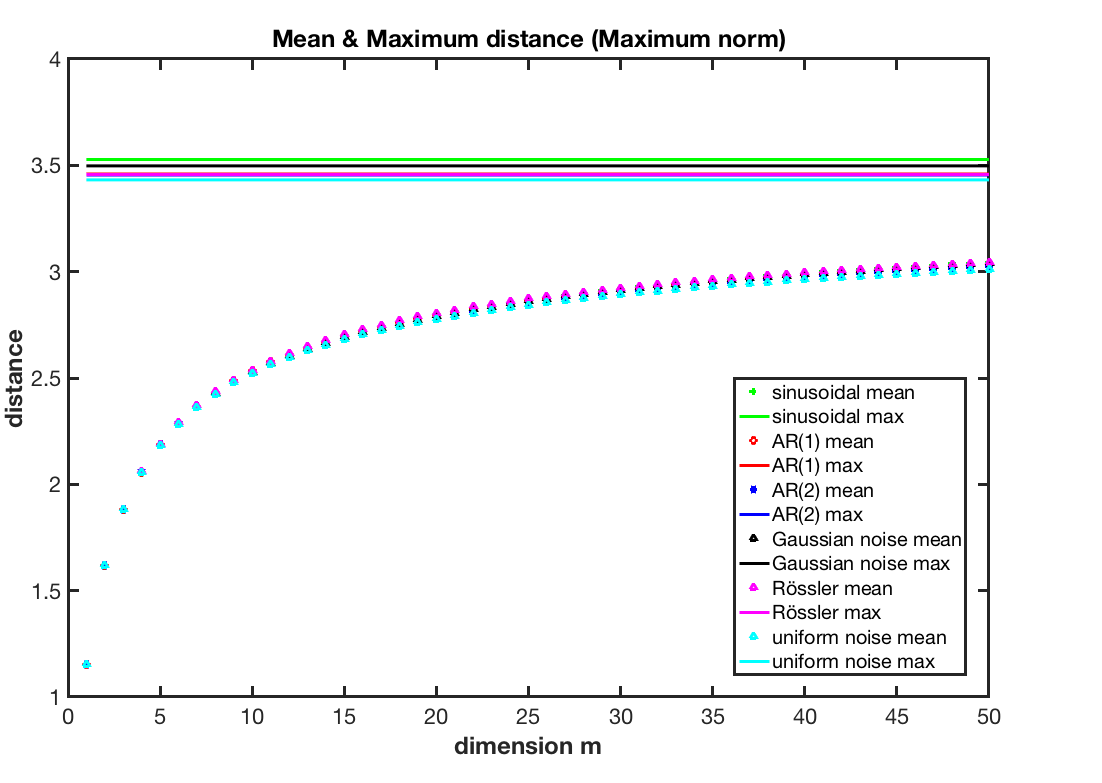}
 \caption{Mean $d^{(\infty)}_{mean}$ and maximum $d^{(\infty)}_{max}$ $L_\infty$ distance between all pairs of state vectors as a function 
of the embedding dimension $m$ for different types of time series: polychromatic harmonic 
oscillation with periods 3, 50 and 500; auto-regressive processes of first and second order 
with parameters $\varphi_1 = 0.5, \varphi_2 = 0.3$; random numbers of standard Gaussian (zero 
mean and unit variance) and uniform (unit variance) distributions, and $x$ component of the 
R\"ossler system (Eq.~\eqref{roessler1}, see Section \ref{numerical}) with $b= 0.3$, $c= 4.5$ and $a$
linearly increasing from 0.32 (spiral chaos) to 0.39 (screw type chaos).}
\label{figure1}
\end{figure} 

Numerical results for different types of systems demonstrate that the largest of all pairwise $L_\infty$ 
distances, $d_{max}^{(\infty)}$, stays constant with increasing embedding dimension, whereas the
mean of all pairwise $L_\infty$ distances, $d_{mean}^{(\infty)}$, monotonically increases with $m$ (Fig.~\ref{figure1}). In order to 
understand this observation, recall that the $L_\infty$ distance between two embedded state 
vectors $\vec{x}_i = (x_{i,1}, x_{i,2},\ldots,x_{i,m})^T$ and $\vec{x}_j = (x_{j,1}, x_{j,2},\ldots,x_{j,m})^T$ is  
\begin{equation}
\|\vec{x}_i-\vec{x}_j\|_\infty = \max_{k=1,\dots,m} \left| x_{i,k}-x_{j,k} \right| = 
d^{(\infty)}_{i,j}(m)
\end{equation}
For $m=1$ (i.e., no embedding), the distance between two observations at times $t_i$ and $t_j$ 
therefore is simply $d^{(\infty)}_{i,j}(1)=\left|x_i-x_j\right|$. For $m=2$, we find
\begin{align}
d^{(\infty)}_{i,j}(2) &= \max \{\left|x_i - x_j\right|,\left|x_{i+\tau}-x_{j+\tau}\right|\} 
\notag \\
&= \max\left\{d^{(\infty)}_{i,j}(1),\left|x_{i+\tau}-x_{j+\tau}\right|\right\} \geqslant 
d^{(\infty)}_{i,j}(1).
\end{align}
By induction, we can easily show that
\begin{equation*}
d^{(\infty)}_{i,j}(m) = \max \left\{ d^{(\infty)}_{i,j}(m-1), \left|x_{i-(m-1)\tau}-x_{j-
(m-1)\tau}\right| \right\} 
\end{equation*}
and therefore
\begin{equation}
d^{(\infty)}_{i,j}(m) \geqslant d^{(\infty)}_{i,j}(m-1) \qquad \forall\ m>1.
\end{equation}
Hence, considering all possible pairs of state vectors $(\vec{x}_i,\vec{x}_j)$ from the time
series, the largest $L_\infty$ distance 
\begin{equation}
d^{(\infty)}_{max}= \max_{i,j}[d^{(\infty)}_{i,j}(1)] = \max_{i,j}[d^{(\infty)}_{i,j}(m)] = d_{max}^{(\infty)} \qquad \forall\ m \notag
\end{equation}
\noindent
cannot change with $m$, since the largest maximum distance will already appear for $m=1$. The mean distance 
\begin{equation}
d_{mean}^{(\infty)}(m) = \frac{1}{N_{\text{eff}}^2(m)} \sum_{i,j=1}^{N_{\text{eff}}(m)} d_{i,j}^{(\infty)}(m), 	\notag
\end{equation}
\noindent
however, necessarily increases with $m$ or stays at most constant. More specifically, as $m$ increases, smaller distances systematically disappear, so that 
the entire distribution is systematically shifted towards its (constant) maximum, thereby 
becoming narrower and exhibiting an increasing mean along with decreasing variance. We conjecture 
that, for large $m$, the distribution of $d^{(\infty)}(m)$ will converge to a limiting 
distribution (see below) possibly depending on the embedding delay $\tau$.

% Section 2B

\subsection{Euclidean norm}\label{influence_euclidean}

\begin{figure}
 \centering
 \includegraphics[scale=0.25]{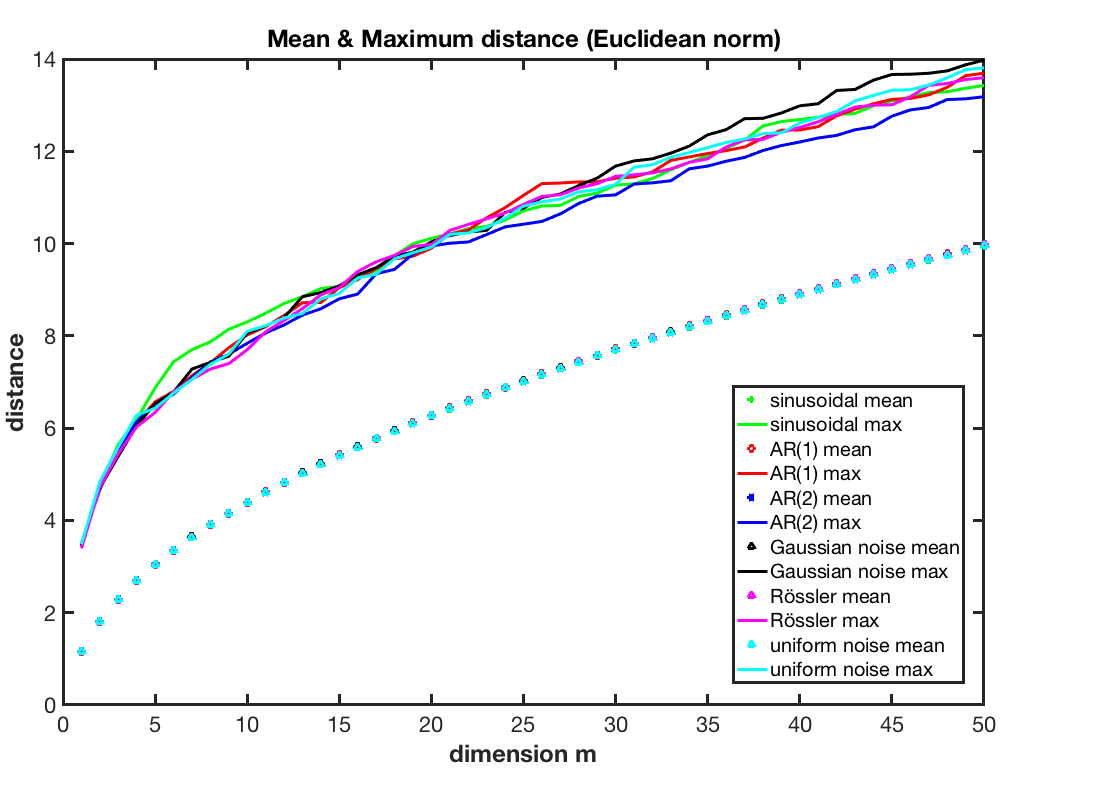}
 \caption{Same as in Fig.~\ref{figure1} for \textit{$L_2$} distances.}
 \label{figure2}
\end{figure}

In case of the \textit{$L_2$} (Euclidean) norm, both mean and maximum of all pairwise distances ($d^{(2)}_{mean}(m)$ and $d^{(2)}_{max}(m)$, respectively) monotonically 
increase with rising $m$ (Fig.~\ref{figure2}). This can be understood as follows: The $L_2$ 
distance between two points in an $m$-dimensional state space, $\vec{x}_i$ and $\vec{x}_j$, is 
given as  
\begin{equation}
\|\vec{x}_i-\vec{x}_j\|_2 = \biggl( \sum_{k=1}^{m}\left|x_{i,k}-x_{j,k}\right|
^2\biggr)^{\frac{1}{2}} = d_{i,j}^{(2)}(m) \label{euc1}
\end{equation}
For the squared $L_2$ distance, this implies:
\begin{align}
\left[d^{(2)}_{i,j}(1)\right]^2 & = (x_i -x_j)^2 \notag \\
\left[d^{(2)}_{i,j}(2)\right]^2 & = (x_i -x_j)^2+(x_{i-\tau} -x_{j-\tau})^2 \notag \\
\label{euc2}&= \left[d^{(2)}_{i,j}(1)\right]^2 +(x_{i-\tau} -x_{j-\tau})^2 \notag \\
&\geqslant \left[d^{(2)}_{i,j}(1)\right]^2 \\
&\vdots \notag\\
\label{euc3} \left[d^{(2)}_{i,j}(m+1)\right]^2&\geqslant \left[d^{(2)}_{i,j}(m)\right]^2 
\geqslant \dots \geqslant \left[d^{(2)}_{i,j}(1)\right]^2,
\end{align}
which explains the observed behavior of both mean and maximum distance using the $L_2$ norm. Specifically, unlike 
for $L_\infty$, the maximum $L_2$ distance between two points is not bound by the largest 
pairwise distance in one dimension. 

In a similar way, we may argue for all $L_p$ distances ($p\in(0,\infty)$) defined as
\begin{equation}
\|\vec{x}_i-\vec{x}_j\|_p = \biggl( \sum_{k=1}^{m}\left|x_{i,k}-x_{j,k}\right|
^p\biggr)^{\frac{1}{p}} = d_{i,j}^{(p)}(m)
\label{Lp_distances}
\end{equation}
that, by the same argument as above,
\begin{equation}
\left[d^{(p)}_{i,j}(m+1)\right]^p \geqslant \left[d^{(p)}_{i,j}(m)\right]^p,
\end{equation}
implying again a monotonic increase of mean and maximum distances with rising embedding dimension (recall the positive semi-definiteness of distances and $p$).

% Section 2C

\subsection{Changing shape of distance distribution with increasing embedding dimension}
\label{influence_shape}

Building upon our previous considerations and numerical results, a mathematically more specific yet challenging question is how exactly an increasing 
embedding dimension $m$ is affecting the shape of the distribution of all pairwise distances 
rather than just its central tendency (mean). 

For the maximum norm, one may argue that the individual components of each embedded state 
vector are commonly constructed such that they are as independent as possible\cite{Fraser}. Accordingly, for 
a system without serial correlations (i.e., uncorrelated noise), the absolute differences 
$d=d^{(\infty)}(1)$ between the components of two state vectors are also independent and 
identically distributed (i.i.d.) and lie within the interval $[0,d_{max}]$.
In 
such case, for sufficiently large $m$, the pairwise $L_\infty$ distance between two state 
vectors can be interpreted as the 
maximum of $m$ i.i.d.~variables that are bounded from above, which should 
follow a reversed Weibull distribution according to the Fisher-Tippett-Gnedenko theorem from extreme value statistics. Note, however, that this expectation is valid only if $m$ is sufficiently large and the i.i.d.~assumption is (approximately) fulfilled, both of which does not 
necessarily have to be the case for real-world time series. Moreover, it is not guaranteed that the given distance distribution in one dimension lies within the domain of attraction of the reversed Weibull class\cite{Leadbetter}, which calls for further theoretical investigation in each specific case.

For other $L_p$ norms including the Euclidean norm, the aforementioned considerations do not apply. For an $L_p$ norm with $p<\infty$,
the pairwise distances $d$ are of the form $(\sum_i z_i^p)^{1/p}$ ($i=1,...,m$) as given in Eq.~\eqref{Lp_distances} with approximately i.i.d.\ variables $z_i$. Hence, the central limit theorem tells us that the distribution of $d^p$ is approximately a normal distribution with mean and standard deviation growing proportionally with $m$ and $\sqrt{m}$, respectively, for large $m$. Therefore, the coefficient of variation of $d^p$ declines approximately as $\sim 1/\sqrt{m}$.
As a consequence, for large $m$ also $d=(d^p)^{1/p}$ is approximately normally distributed with mean and standard deviation growing approximately as $\sim m^{1/p}$ and $\sim\sqrt{m}\frac{dz^{1/p}}{dz}|_{z=m}\sim\sqrt{m}m^{1/p - 1}=m^{1/p - 1/2}$. The coefficient of variation of $d$ thus behaves approximately as $\sim m^{1/p - 1/2}/m^{1/p} = 1/\sqrt{m}$, just as for $d^p$. In other words, the relative variability of $d$ narrows in the same fashion for all $p<\infty$ as $m$ grows, and only the growth of the absolute scale of $d$ with $m$ depends on $p$, which is also known as the ``curse of dimensionality''\cite{Zimek}. 

\begin{figure*}
 \centering
 \includegraphics[scale=0.28]{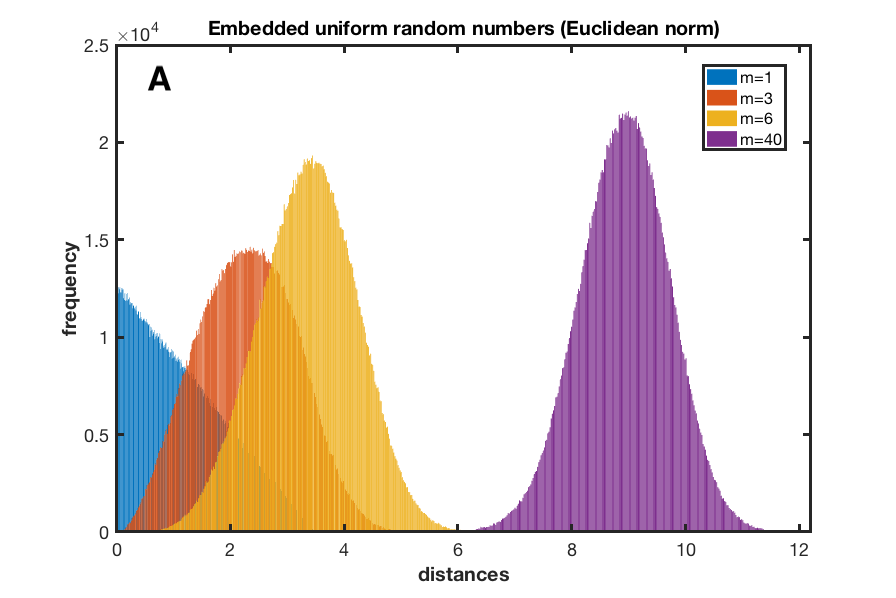} \hfill
 \includegraphics[scale=0.28]{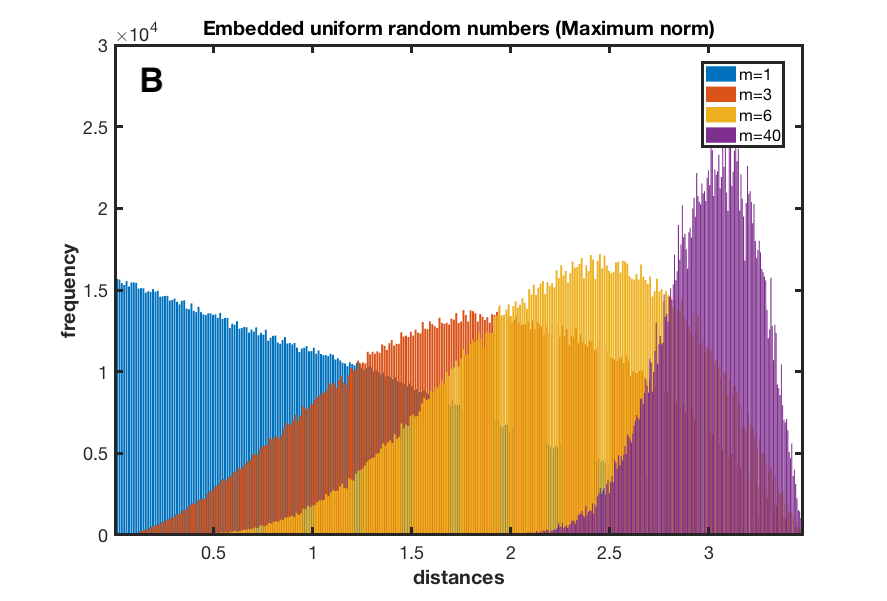} \\
 \includegraphics[scale=0.28]{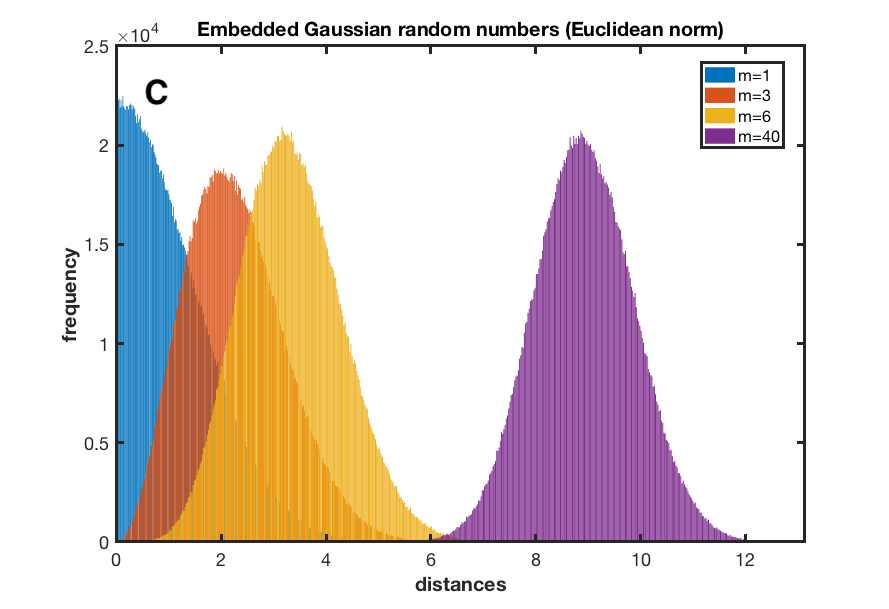} \hfill
 \includegraphics[scale=0.28]{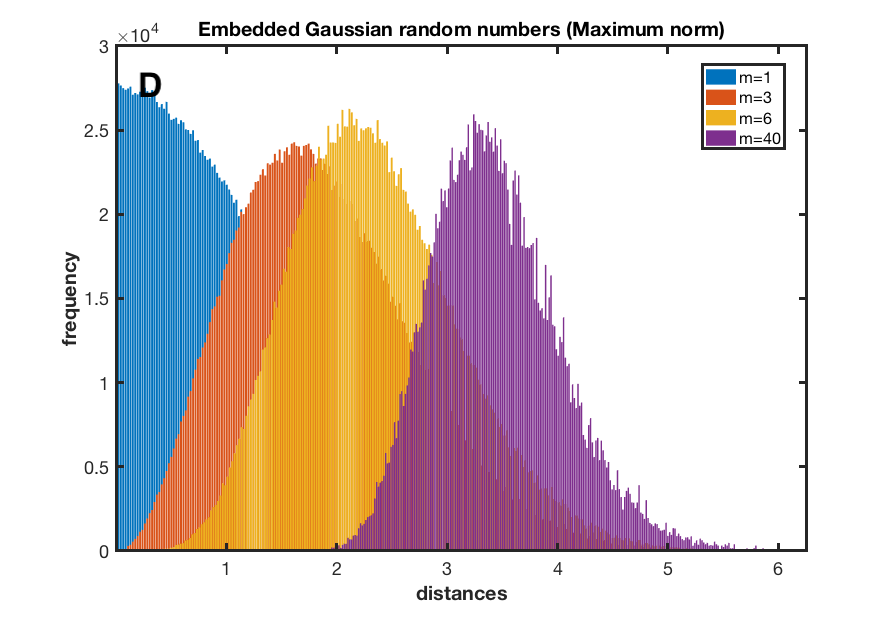} \\
 \includegraphics[scale=0.28]{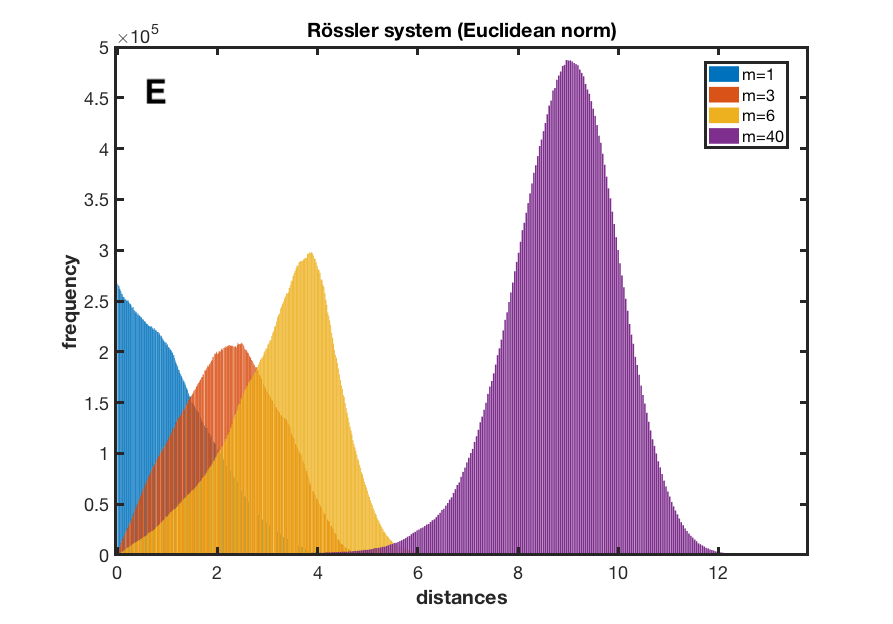} \hfill
 \includegraphics[scale=0.27]{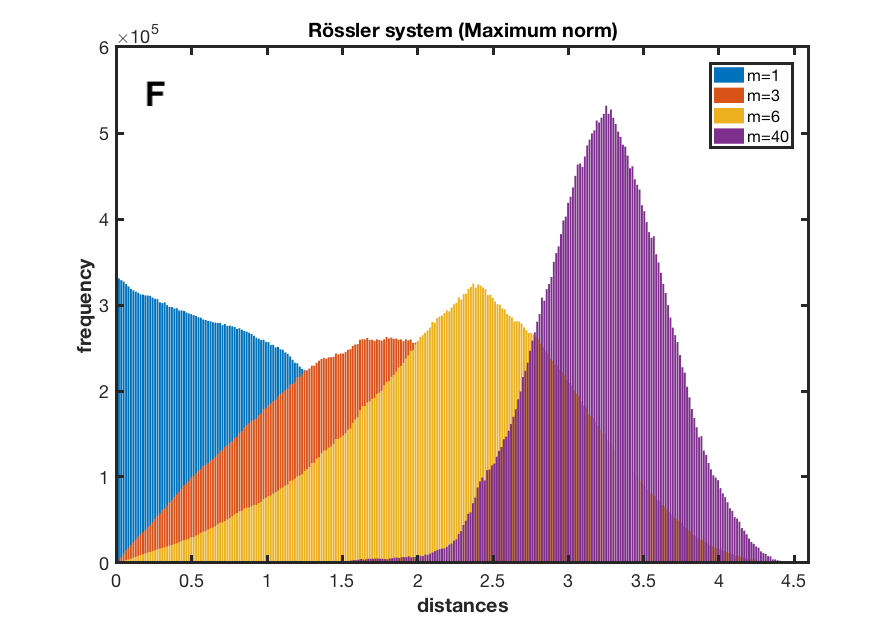} \\
\caption{Selected histograms of the $L_2$ (A,C,E) and $L_\infty$ (B,D,E) distances of $N=1,500$ independent random numbers with uniform (A,B) and Gaussian (C,D) distribution as well as (E,F) for the $x$ component of the R\"ossler system (Eq.~\eqref{roessler1}, $N=6,000$, see Section \ref{numerical}) with control parameters $b=0.3$, $c=4.5$ and $a$ linearly increasing from 0.32 (spiral chaos) to 
0.39 (screw type chaos) for different embedding dimensions $m$.}
\label{dist_examples}
\end{figure*}

The considerations made above do explain the numerical results in
Fig.~\ref{dist_examples}, showing histograms of the distances of three different time series for selected values of the embedding dimension $m$ and for the $L_2$ and $L_{\infty}$ norms. In addition to time series fulfilling the i.i.d.~assumption (Fig.~\ref{dist_examples} A,B,C,D), here we are also interested in deterministic systems. As an illustrative example, we choose here the R\"ossler system (Eq.~\eqref{roessler1}, Fig.~\ref{dist_examples} E,F) in some non-stationary (drifting parameter) setting, which will be further studied in Section \ref{numerical}.

In this regard, it is confirmed that the expectation value of the distance distribution takes higher values with increasing $m$. The probability to find small distances therefore decreases. In case of the $L_{\infty}$ norm (Fig.~\ref{dist_examples} B,D,F), this growth is bounded and we can identify a convergence of the distribution, in some cases eventually towards the aforementioned reversed Weibull distribution. In turn, for the $L_2$ norm (Fig.~\ref{dist_examples} A,C,E) the convergence towards a normal distribution is discernible. Considering the R\"ossler time series (Fig.~\ref{dist_examples} E,F), the empirical expectations are approximately met by the observations, even though the distribution of $L_\infty$ distances exhibits a slightly more complex (i.e., less symmetric) shape than for the two noise series. Specifically, for the $L_2$ norm the resulting distance distribution is left-skewed with a pronounced lower tail (see Fig.~\ref{dist_examples} E), whereas for the $L_\infty$ norm we observe a more triangular rather than Weibull-like shape. Notably, the i.i.d.~assumption is violated when dealing with such a deterministic dynamical system, even though we have chosen the time delay $\tau$ such as to minimize serial correlations (see Section~\ref{numerical}). 

\begin{figure*}
 \centering
 \includegraphics[scale=1.935]{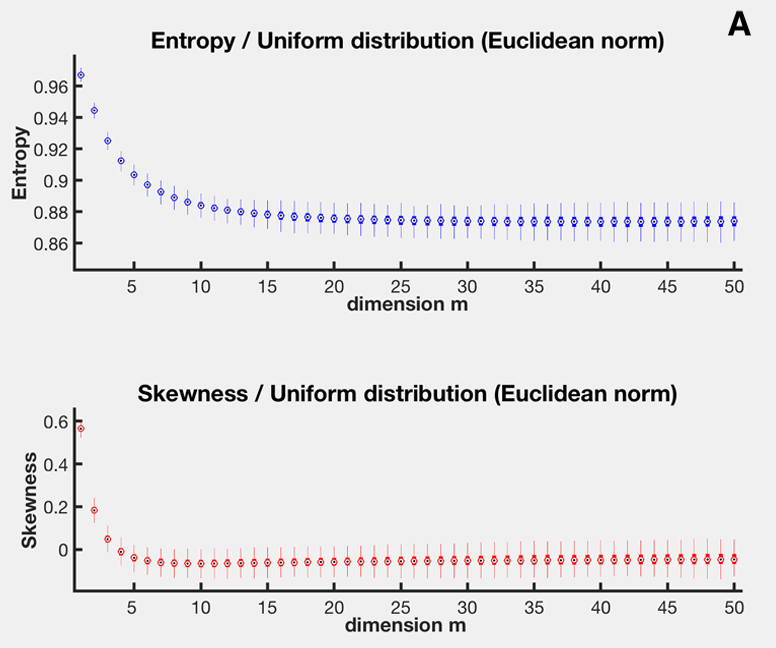} \hfill
 \includegraphics[scale=1.935]{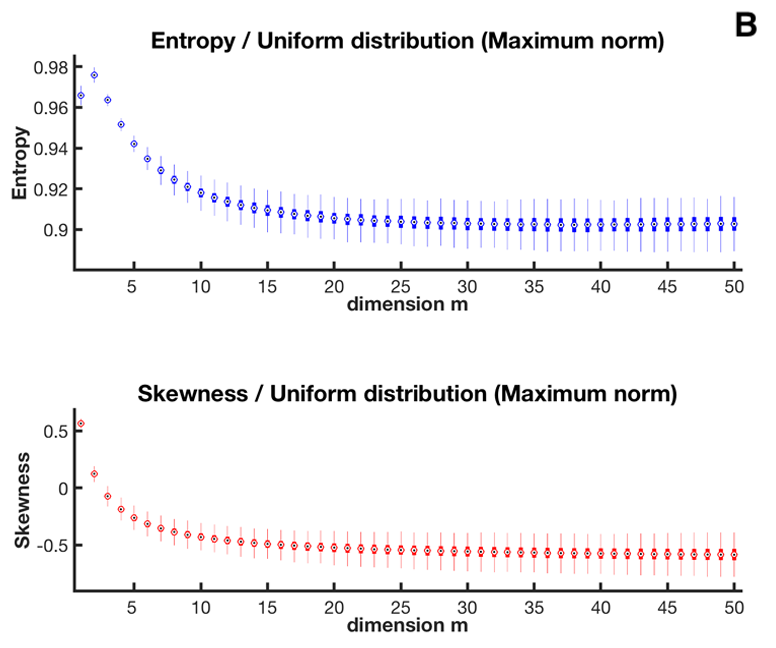} \\
 \includegraphics[scale=1.935]{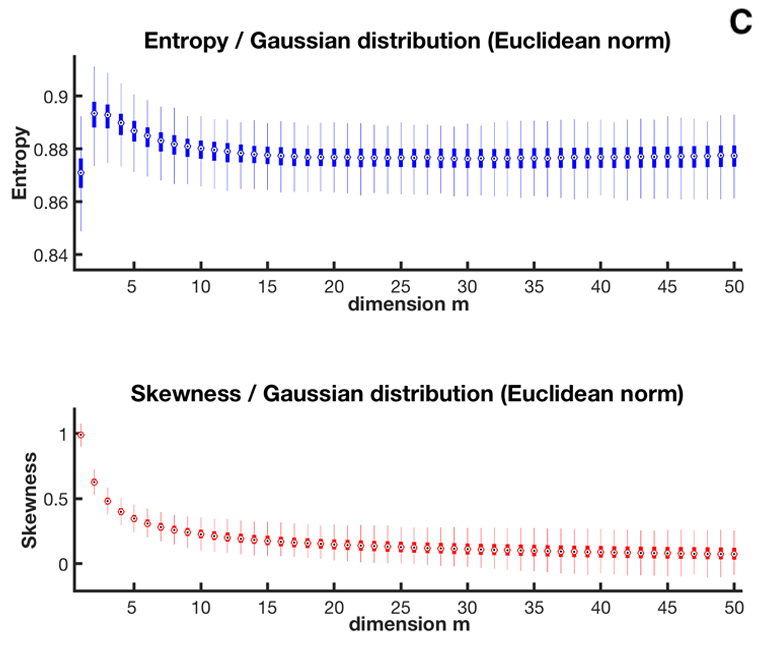} \hfill
 \includegraphics[scale=1.935]{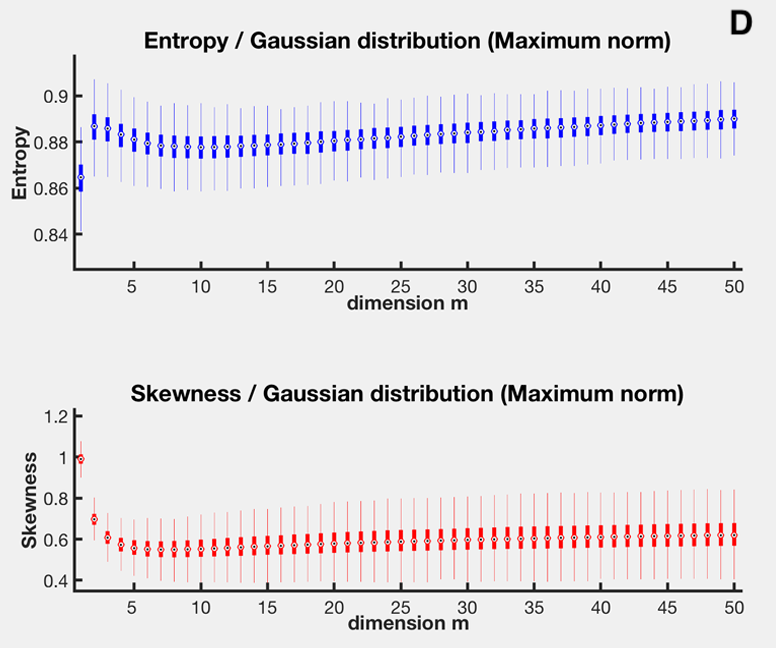} \\
 \includegraphics[scale=0.27]{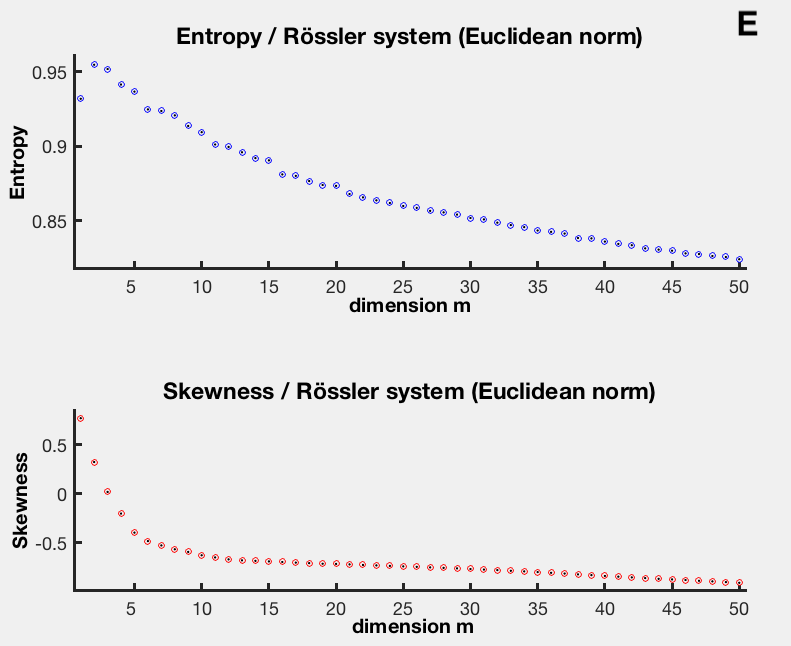} \hfill
 \includegraphics[scale=0.27]{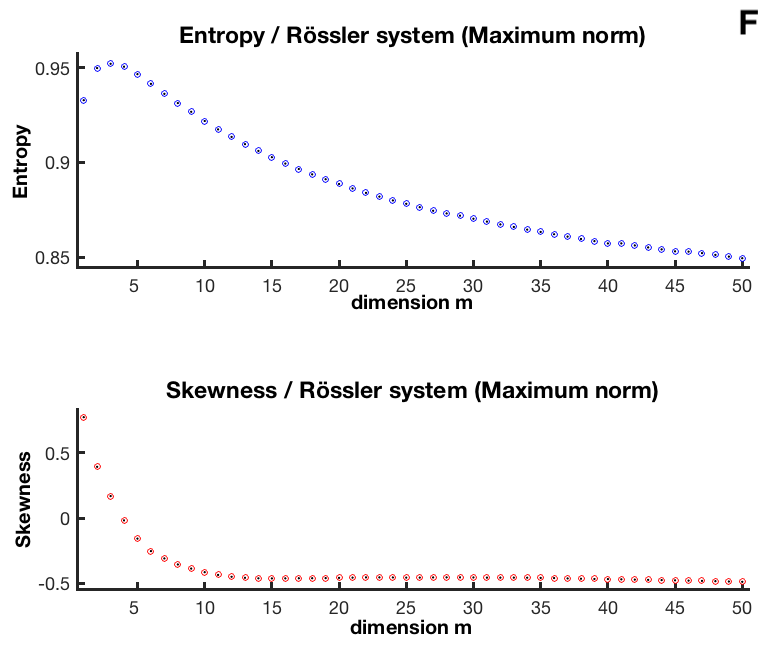}
\caption{Skewness (red) and Shannon entropy (blue) of the $L_2$ (A,C,E) and $L_\infty$ (B,D,E) distances of $N=1,500$ independent random numbers with uniform (A,B) and Gaussian (C,D) distribution and (E,F) the $x$ component of the R\"ossler system (Eq.~\eqref{roessler1}, $N=6,000$, see Section \ref{numerical}) with control parameters $b=0.3$, $c=4.5$ and $a$ linearly increasing from 0.32 (spiral type chaos) to 0.39 (screw type chaos) as a function of the embedding dimension $m$. For the two noise series, box plots show the variability estimated from 1,000 independent realizations for each data set.}
\label{statistics}
\end{figure*}

In order to further characterize the shape of the empirically observed pairwise distance distributions shown in Fig.~\ref{dist_examples}
in more detail, we consider two standard characteristics from descriptive statistics. On the one hand, the skewness
\begin{equation}
\hat{s} = \frac{\frac{1}{N_d}\sum_{i=1}^{N_d}(d_i-\bar{d})^3}
{\left( \sqrt{\frac{1}{N_d}\sum_{i=1}^{N_d}
 (d_i - \bar{d})^2} \right)^3} \label{skewness}
\end{equation}
of the distribution measures its asymmetry around the sample mean distance \textit{$\bar{d}$}.
On the other hand, we study the associated Shannon entropy
\begin{equation}
\label{shannon} \hat{h} = -\sum_{j=1}^{N_b} p_j
\frac{\log(p_j)}{\log(N_b)}
\end{equation}
providing an integral measure of the heterogeneity of the distribution of $d$. Here, $j$ enumerates the bins of a histogram of the values of $d$ with $N_b$ bins and relative frequencies $p_j$, and $N_d$ is 
the number of pairwise distances in the sample (i.e., the number of independent entries of the 
distance matrix $\mathbf{d}$, $N_d=N_\text{eff}(N_\text{eff}-1)/2$). 
The bin width has been selected by first computing the optimum value according to the Freedman-Diaconis rule\cite{Freedman} for each embedding dimension $m$ and then averaging over all corresponding values and taking the resulting mean to keep $N_b$ fixed for each considered setting. Specifically, for the time series drawn from the Gaussian and uniform distributions in Figs.~\ref{dist_examples} and \ref{statistics}, $N_{b,L_2}=355$ and $N_{b,L_{\infty}}=286$, while for the R\"ossler system,  $N_{b,L_2}=701$ and $N_{b,L_{\infty}}=771$.

According to the corresponding normalization, $\hat{h}$ 
assumes its maximum of one in case of a uniform distribution (since then, $p_j=1/N_b \quad \forall\, j=1,...,N_b$, i.e., for each distance out of $[d_{min},d_{max}]$). In turn, the more heterogeneous (e.g., spiky or generally asymmetric) the distribution of distances gets, the lower $\hat{h}$.

Figure~\ref{statistics} shows the resulting behavior of both characteristics for the
$L_2$ (panels A,C,E) and $L_\infty$ (panels B,D,F) distances obtained from uniform and
Gaussian distributed noise as well as for the non-stationary R\"ossler system (Eq.~\eqref{roessler1},
see Section~\ref{numerical}) in dependence on the embedding dimension. The results
complement the qualitative description based on a visual inspection of Fig.~\ref{dist_examples} as given above. In case of the
$L_2$ norm and time series drawn from uniform and Gaussian distributions (Fig.~\ref{statistics} A,C) we observe the skewness converging towards zero (symmetric Gaussian
distribution) and the entropy reflecting this convergence towards a normal distribution
by a downward trend until the skewness approaches zero as $m$ further increases.
Although the theoretically predicted Gaussian shape for high $m$ is visually apparent
in case of the time series from the R\"ossler system (see Fig.~\ref{dist_examples} E,), the skewness takes clearly non-zero negative values while the entropy constantly decreases with increasing $m$, indicating an asymmetric shape (Fig.~\ref{statistics} E). In case of the $L_\infty$ norm, the results show the expected convergence of the distance distribution to a limit distribution with non-zero skewness as typical for extreme value distributions. 

Based on the above findings, we emphasize that it is not straightforward to analytically describe the shape of the distance distribution of an embedded time series stemming from an arbitrary dynamical system with potentially nontrivial serial correlations. Regarding the question how we could automatically choose a recurrence threshold independent of the embedding dimension, the chosen norm and the underlying system, we point out that the problem is neither the general increase nor the successive concentration of distances when increasing the embedding dimension. These factors could be easily accounted for by relating the threshold selection to the spatial extend of the state space object, similar as, for instance, suggested by Abarbanel\cite{Abarbanel} in the context of the false nearest neighbor algorithm. What is more, our findings suggest that taking the varying shape of the distance distribution at increasing embedding dimensions into consideration is key to an appropriate, embedding dimension independent recurrence threshold estimation method. Thus we recommend to use a numerical estimate of a certain (sufficiently low) percentile of the distance distribution as threshold rather than a percentage of the maximum or mean state space diameter. This leads to a constant global recurrence rate (which equals the chosen percentile) and ensures the invariance of recurrence statistics under arbitrary embedding dimensions and the chosen norm, as we will exemplify in the following section.    

Even more, by conserving the recurrence rate, possible dependences of RQA characteristics on the density of recurrences for different $m$ are omitted, and corresponding residual changes of these measures upon varying $m$ may rather point to either insufficiently low embedding dimension (missing essential factors contributing to the system's dynamics, in a similar spirit as, e.g., for the false nearest neighbor method) or spurious recurrence structures arising from overembedding\cite{Thiel2006}. These ideas should be further studied in future work.

% Section 3

\section{Numerical example}\label{numerical}

In this section, we will demonstrate the effect of the varying shape of the distance distribution with increasing embedding dimension on different threshold selection approaches working with a globally fixed value of $\varepsilon$. Specifically, we will study the behavior of a few selected RQA and recurrence network characteristics resulting from a low dimensional embedded time series and compare these to the results obtained using a higher-dimensional embedding. We focus on the question how the different threshold estimation methods 
qualitatively change the RQA characteristics with increasing embedding dimension (rather than the question what these quantities effectively measure in terms of specific nonlinear dynamical properties of the system under study).
 
In order to mimic a practically relevant test case of a non-stationary low-dimensional dynamical system, where we should use some higher embedding dimension (following \citeauthor{Heg}\citep{Heg}) instead of a moderate choice, we consider the classical R\"ossler system\cite{Roe}
\begin{equation}
\begin{array}{rcl}
\dot{x}&=&-y-z \\
\dot{y}&=&x+ay \\
\dot{z}&=&b+z(x-c).
\end{array}
%\notag \dot{x} &= -y -z \\
%\notag \dot{y} &= x+ay\\
\label{roessler1}
%\dot{z} &= b+z(x-c)
\end{equation}
For numerically solving this system of equations, we use a fourth-order Runge-Kutta integrator 
with the initial conditions $x(0)=1$, $y(0)=1$ and $z(0)=0$, an integration step of $t_\text{is} 
= 0.001$ and a total of 1,300,000 iterations. Therefore, we simulate the system's evolution over 1,300 time units (t.u.). By using a sampling interval of $\delta t = 0.2$ t.u.~we obtain 6,500 samples forming our time series for the three components $x$, $y$ and $z$.
We remove the first 500 points ($\widehat{=}100$~t.u.) that could be affected by transient dynamics and 
retain the remaining 6,000 points ($\widehat{=}1200$~t.u.) of the $x$ component for further analysis.

Depending on the parameters $a$, $b$ and $c$, the system (Eq.~\eqref{roessler1}) exhibits either 
regular or chaotic dynamics. Here, we consider a transitory setting previously studied by \citeauthor{Mal1}\cite{Mal1}, where the parameter $a$ increases from 0.32 to 0.39 while 
keeping $b=0.3$ and $c=4.5$ fixed. In this case, the system undergoes a transition from spiral 
to screw type chaos, which originates from the presence of a homoclinic
orbit\cite{Gas} and results in a transient, quasi-periodic phase as $a$ rises. Note that instead of studying the stationary R\"ossler system for different values of $a$ as done in previous studies\cite{Zou2012}, we intentionally employ a gradual parameter change leading to a non-stationary system which calls for 
a systematic overembedding when performing nonlinear time series analysis\citep{Heg}. 
Specifically, we implement a linear variation of $a$ as
\begin{equation}
a(t) = a_1 + 5.83\cdot 10^{-8}t.
\end{equation} 

For the resulting time series, we now study the associated RPs together with the time-dependence of a few selected properties of RQA and recurrence network analysis. For the latter purpose, we use a running window over the (global) RP with a window size of $w=400$ and mutual 
shift of $ws=40$ data points, i.e., 90\% overlap between consecutive windows. Since we are 
aiming to study the change of recurrence properties associated with a chaos-chaos transition, 
we choose a set of three characteristics that have been previously shown to sensitively trace 
the associated variations\cite{Marwan2,Marwan3}:

\begin{itemize}
\item the \textit{recurrence rate}\cite{Marwan1}
\begin{align}
\label{RR} RR(\varepsilon) = \frac{1}{N^2} \sum_{i,j=1}^N R_{i,j}(\varepsilon),
\end{align}
\item the \textit{laminarity}\cite{Marwan1,Marwan2} 
\begin{align}
\label{LAM}LAM = \frac{\sum_{v=v_{\min}}^N v p(v)}{\sum_{v=1}^N v p(v)},
\end{align}
where $p(v)$ is the relative frequency of vertical line structures of length $v$ in the RP,
\begin{align}
\notag p(v) = \sum_{i,j=1}^N (1-R_{i,j})(1-R_{i,j+v}) \prod_{k=0}^{v-1} R_{i,j+k},
\end{align}
and the minimal vertical line length considered is set to $v_\text{min}=4$, and
\item the \textit{transitivity}\cite{Marwan3,Don}
\begin{align}
\label{TRANS}\mathcal{T} = \frac{\sum_{i,j,k=1}^N A_{j,k}A_{i,j}A_{i,k}}{\sum_{i,j,k=1}^N 
A_{i,j}A_{i,k} (1-\delta_{j,k})}
\end{align}
with $A_{i,j}=R_{i,j}-\delta_{i,j}$.
\end{itemize}
\noindent
Note that even in settings where we have fixed the global recurrence rate of the full RP, local variations of this property (as well as the two other characteristics) may arise die to the non-stationarity of the considered system.

\begin{figure*}
 \centering
 \includegraphics[width=\textwidth]{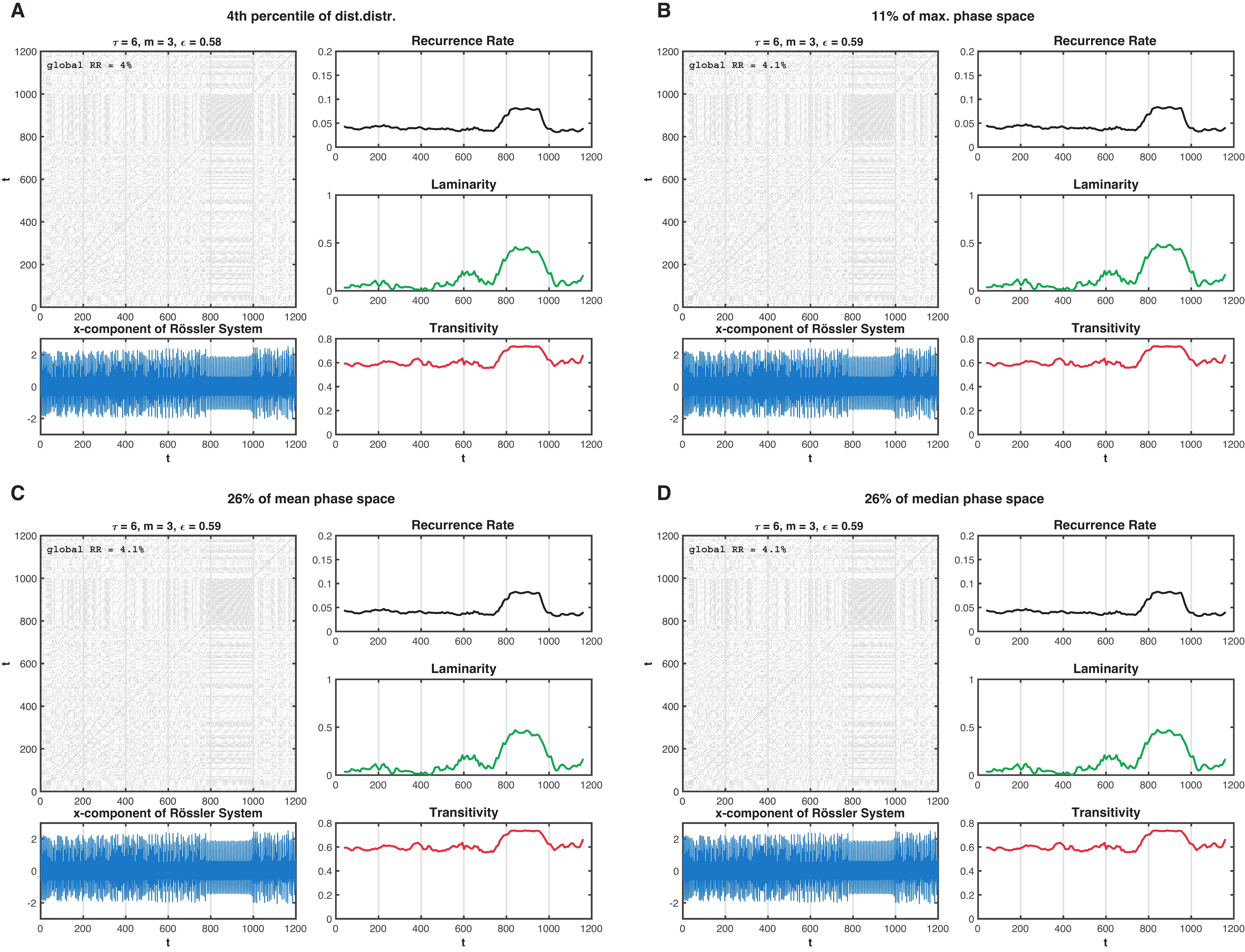}
  \caption{RPs and time-dependent recurrence characteristics $RR$ (black), $LAM$ (green) and $
\mathcal{T}$ (red) based on the $x$ component of the non-stationary R\"ossler system (see text 
for details), using the $L_2$ norm. Shown are the results for low-dimensional embedding ($m=3$) and for four different methods to select the recurrence threshold according to a certain percentile of the distance distribution (A) and according to some percentage of the maximum (B), mean (C) or median distance (D) of state vectors on the reconstructed attractor. The actual values ($4$th percentile, $11\%$, $26\%$ and $26\%$, respectively) are chosen such that the global recurrence rate for each method is $\approx 4\%$.}
\label{results_roessler3}
\end{figure*} 

The RPs and the associated time-dependent recurrence characteristics for a ``normal'' three-dimensional embedding of the non-stationary 
R\"ossler system with embedding delay $\tau=6$, consistent with the first local minimum of the mutual information\cite{Fraser}, are shown in Fig.~\ref{results_roessler3} using the Euclidean norm. We compare the results for four different choices of a fixed recurrence threshold, which are expected to give reasonable results in arbitrary embedding dimensions. Panel A shows the proposed method of taking a certain percentile of the distance distribution, while panels B, C and D correspond to a threshold selection according to some percentage of the maximum, mean and median distance of state vectors on the attractor in the reconstructed state space. We choose the actual value for each method such that a global recurrence rate of $RR\approx 4\%$ is achieved in all four cases. Comparing the different panels, as expected there are hardly any marked differences in the RP or the temporal 
changes of the recurrence characteristics, because we are dealing with a setting resulting in (more or less) the same global recurrence rate. The transition from spiral to screw type chaos (see above) between $t_1\approx 780$ and $t_2\approx 1000$ is well resolved by all three measures $RR$, $LAM$ and $\mathcal{T}$. Note that the chosen RQA characteristics principally detect the temporary quasi-periodic regime rather than showing significantly different levels for the distinct chaotic attractors. However, using a different setup of numerical experiments, it could be possible to also detect quantitative differences between the recurrence properties of the two distinct chaotic attractors. Specifically, Zou \textit{et~al.}\cite{Zou2012} found different levels of $\mathcal{T}$ in stationary simulations of the R\"ossler system with different, constant values of $a$ and $\varepsilon$ chosen such as to maintain the same recurrence rate in all settings. Further corresponding analyses are outlined as a subject of some separate study.

\begin{figure*}
 \centering
 \includegraphics[width=\textwidth]{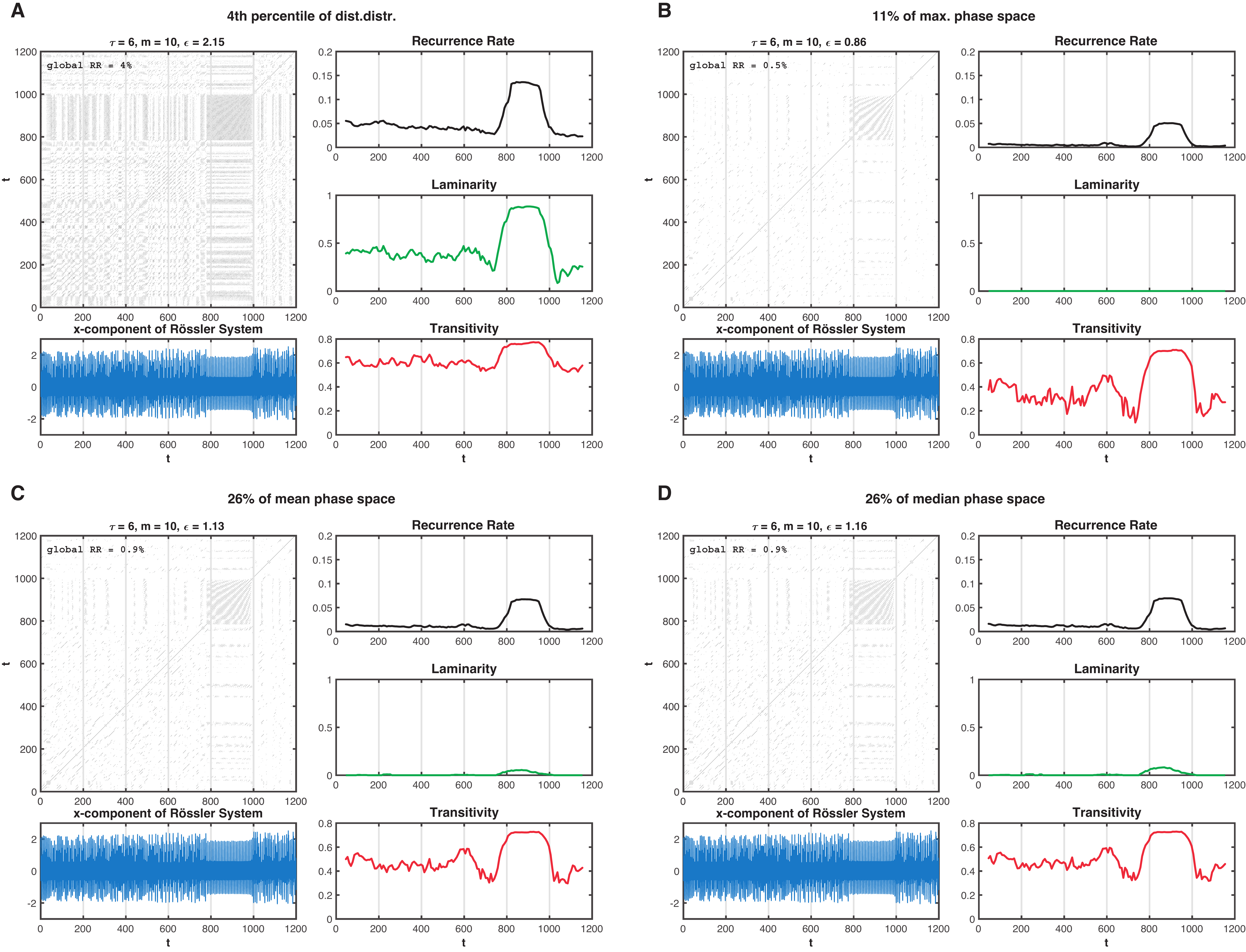}
  \caption{Same as in Fig.~\ref{results_roessler3} for higher-dimensional embedding ($m=10$). 
  %RPs and time-dependent recurrence characteristics $RR$ (black), $LAM$ (green) and $\mathcal{T}$ (red) based on the $x$ component of the non-stationary R\"ossler system (see text for details), using the $L_2$ norm like in Fig.~\ref{results_roessler3}. Here the results for high-dimensional embedding ($m=10$) and for the four different methods to calculate the recurrence threshold are shown. 
Note the drop in the global recurrence rate in panels B, C, D in comparison to A or Fig.~\ref{results_roessler3} and the arising change in the values of all recurrence characteristics.}
\label{results_roessler10}
\end{figure*} 

However, if choosing a higher-dimensional embedding ($m=10$) motivated by the 
non-stationarity of the system, the RP becomes almost completely white if the recurrence threshold 
is chosen based upon the same percentages of the maximum, mean or median state space distances as used before (Fig.~\ref{results_roessler10}B,C,D)
In this case, $RR$ exhibits values near zero except for the transitional phase. The 
transitivity $\mathcal{T}$ is also still able to detect the transition, whereas $LAM$ completely fails in this task. 
In contrast, we retain the same density of recurrences and, hence, resolution of the RP as for $m=3$ when fixing the threshold according to the whole distance distribution (Fig.~\ref{results_roessler10}A). Here, the chaotic dynamics and the chaos-chaos transition are well recovered by all three measures. Considering the results of Section \ref{influence}, the reason for the failure of the methods based on individual statistical properties (maximum, mean, median) of the pairwise distance distribution between state vectors is the change in the shape of that distribution beyond its characteristic location and range parameters. Hence, we argue that selecting the recurrence threshold at some percentile of the distance distribution is to be preferred if we aim to obtain reliable results for a broad range of embedding dimensions, which is the case if we wish to automatically choose fixed recurrence thresholds for the analysis of arbitrary complex systems.

% Section 4

\section{Conclusions}\label{conclusion}

We have presented both empirical and numerical results concerning an appropriate strategy to choose the recurrence threshold in applications of recurrence plots and quantitative analyses of their properties. In this context, we have provided arguments in support of selecting a fixed recurrence rate according to some (lower) precentile of the pairwise distance distribution of state vectors, which is to be chosen in practice according to the available time series length. Our findings suggest that the recurrence characteristics obtained when applying this strategy are largely independent of the chosen norm and embedding dimension. In turn, we have shown
that the latter is not the case when selecting the recurrence threshold according to a certain percentage of the mean or maximum state space diameter, as sometimes suggested in other works\cite{Zbi1,schinkel2008}. It also suggests that alternative approaches, such as normalizing the time series and applying a uniform threshold independent of the embedding dimension and the considered norm\cite{Jacob2016PRE}, are not likely to perform well for any kind of data, when neglecting the effect on the distance distribution with increasing embedding dimension.

At the conceptual level, it appears beneficial to select a recurrence threshold according to the specific topological and geometric properties of the observed trajectory in the (reconstructed) state space. In support of our general considerations, we have qualitatively discussed the behavior of the distribution of distances in state space along with some quantitative results, leading to some interesting questions associated with the convergence properties of these distributions at high embedding dimensions that should be further addressed in future studies. Notably, the relationship between the distribution of $L_\infty$ distances and extreme value statistics clearly deserves further investigations to fully understand the emerging shape of the distributions as the embedding dimension becomes large.

As a cautionary note, we would like to stress that the considerations presented in this work relate exclusively to the concept of time-delay embedding as the most widely applied embedding technique in the context of state space reconstruction associated with nonlinear time series analysis. However, there exist alternative embedding techniques such as derivative embedding\cite{Lekscha}, for which the metric properties of different components of the embedding vector cannot be easily related to each other. In such situations, we expect the monotonic rise in the mean distances among all pairs of state vectors with increasing embedding dimension to persist as in the case of time-delay embedding discussed here, whereas the actual convergence properties and shape of the limit distributions might be essentially different and present an open subject for further research.

Taken together, we are confident that the results presented in this work provide an important step towards automatizing the problem of selecting recurrence thresholds in some data-adaptive way. The latter is key for further widening the scope of applications of recurrence plots, recurrence quantification analysis and related techniques across scientific disciplines. Especially in the context of long time series originating from non-stationary systems, which frequently appear in many fields including Earth and life sciences, having a generally applicable approach at hand is crucial for obtaining reliable and easily interpretable results.

\section*{Acknowledgments}
This work has been financially supported by the German Research Foundation (DFG project no. MA 4759/9-1 and MA4759/8), the European Union's Horizon 2020 Research and Innovation Programme under the Marie Sklodowska-Curie grant agreement no.~691037 (project QUEST), and the German Federal Ministry of Education and Research via the Young Investigators Group CoSy-CC$^2$ (Complex Systems Approaches to Understanding Causes and Consequences of Past, Present and Future Climate Change, grant no. 01LN1306A).

\bibliography{refs_paper}

\end{document}